\documentclass[10pt,letterpaper,twocolumn,prl,showpacs]{revtex4-1}

\usepackage[english]{babel}

\usepackage{amsmath,amssymb}
\usepackage[dvips]{graphicx,color}
\usepackage[colorlinks=true, allcolors=blue]{hyperref}

\begin{document}

\title{Widely tunable optical parametric oscillation in a Kerr microresonator}

\author{Noel~Lito~B.~Sayson, Karen~E.~Webb$^*$, St\'ephane~Coen, Miro~Erkintalo, and Stuart~G.~Murdoch}

\affiliation{The Dodd-Walls Centre for Photonic and Quantum Technologies, Department of Physics, The University of Auckland, Private Bag 92019, Auckland 1142, New Zealand \\
$^*$Corresponding author: s.murdoch@auckland.ac.nz }

\begin{abstract}
We report on the first experimental demonstration of widely-tunable parametric sideband generation in a Kerr microresonator. Specifically, by pumping a silica microsphere in the normal dispersion regime, we achieve the generation of phase-matched four-wave mixing sidebands at large frequency detunings from the pump. Thanks to the role of higher-order dispersion in enabling phase matching, small variations of the pump wavelength translate into very large and controllable changes in the wavelengths of the generated sidebands: we experimentally demonstrate over 720~nm of tunability using a low-power continuous-wave pump laser in the C-band. We also derive simple theoretical predictions for the phase-matched sideband frequencies, and discuss the predictions in light of the discrete cavity resonance frequencies. Our experimentally measured sideband wavelengths are in very good agreement with theoretical predictions obtained from our simple phase matching analysis.
\end{abstract}

\maketitle

\noindent The ability to generate coherent light that is widely tunable in wavelength is highly desirable for numerous applications \cite{demtroder,duarte}. Whilst optical parametric oscillators based on second-order $\chi^{(2)}$ nonlinear crystals allow for generation of laser-like light at virtually any wavelength \cite{savage10}, their widespread adoption is hindered by the difficulty of realising devices that are compact and low-cost. An alternative solution is to generate large frequency shift parametric sidebands via degenerate four-wave mixing (FWM) in optical fibers through the $\chi^{(3)}$ Kerr nonlinearity. Such sideband generation was first reported in single-pass transmission experiments~\cite{stolen75,stolen81,lin81b,lin81}, and a variety of modal~\cite{stolen75,stolen81,lin81b} and dispersive~\cite{lin81,harvey03,pitois03,marhic04} phase matching schemes have been demonstrated.

In single-mode optical fibers (or other Kerr nonlinear waveguides), large frequency shift parametric sidebands can be generated via FWM when pumping in the regime of normal group-velocity dispersion \cite{harvey03,pitois03}. In this regime, phase matching can be achieved through higher-order dispersion, and small changes of the pump wavelength can translate into very large changes in the generated sideband frequencies. To enhance the efficiency of sideband generation, \emph{fiber optical parametric oscillators} have been demonstrated that exploit these ideas~\cite{knox,wong07,sharping,bessin17}. Such devices have reached very impressive sideband tuning ranges of up to an octave~\cite{sharping07,xu08}. Yet, the low finesse of fiber-based devices necessitates the use of high-power (often pulsed) pump fields to achieve parametric oscillation, limiting the devices' practical usefulness. Very recently, a new candidate platform has emerged with the potential to enable efficient generation of large frequency shift FWM sidebands using low-power continuous wave (CW) lasers: the high-Q optical microresonator~\cite{liang15,matsko16,fujii17,huang17}. The extraordinarily high finesse of these devices can dramatically reduce the threshold for nonlinear effects \cite{kippenberg04}, raising the intriguing possibility of low-power and low-cost sources of widely tunable light.

Whilst large frequency shift parametric oscillation has already been observed in microresonators, an experimental demonstration of their full tuning capabilities has remained elusive. Fixed frequency parametric sidebands, phase-matched through higher-order dispersion and detuned by about 40~THz from the pump, have been observed in both MgF$_2$ microdisks \cite{matsko16} and silica microtoroids \cite{fujii17}. Even larger fixed frequency shifts (up to 74.5~THz) have been demonstrated in MgF$_2$ microdisks by exploiting modal phase matching~\cite{liang15}. Most recently, a novel quasi-phase matching scheme was implemented in a Si$_3$N$_4$ ring resonator to produce a pair of parametric sidebands detuned by 40~THz from the pump that could be discretely tuned over a small range ($\sim1.5$~THz) as the pump wavelength was varied by a similar amount ($\sim1$~THz)~\cite{huang17}. So far, however, no experimental demonstrations of \emph{wide} tunability has been reported.

In this Letter, we experimentally demonstrate widely-tunable parametric oscillation -- phase-matched through higher-order dispersion -- in a Kerr microresonator. Specifically, by driving a silica microsphere with a low-power CW pump source in the C-band, we are able to realize narrowband parametric sidebands with a total tuning range of 720~nm (85~THz). The observed sideband wavelengths are in excellent agreement with simple phase matching calculations that account for higher-order dispersion. Our results, achieved with only 50~mW of pump power, clearly demonstrate the potential of $\chi^{(3)}$ microresonators to operate as compact, low-power, tunable optical sources.

We begin by presenting a theoretical analysis of large frequency shift sideband generation in a CW driven Kerr resonator. We are interested in the ``scalar'' regime where the pump, signal, and idler waves all belong to the same mode family. In this case, the evolution of the intracavity field $E(t,\tau)$ obeys the generalized mean-field Lugiato-Lefever equation (LLE)~\cite{coen13a}:
 \begin{equation}
\begin{split}
t_\textrm{R}\frac{\partial E(t,\tau)}{\partial t}=&\left[-\alpha-i\delta_0+i\sum_{k\geq2}\frac{\beta_k}{k!}\left(i\frac{\partial}{\partial\tau}\right)^k\right]E\\
&+i\gamma|E|^2E+\sqrt{\theta}E_\textrm{in}.
\end{split}
\label{lle}
\end{equation}
Here, $t$~is a slow time variable that describes the evolution of the field envelope over successive round trips, while $\tau$~is a fast time variable, defined in a co-moving reference frame, that allows for the description of the envelope's profile over a single round trip. $t_\textrm{R}$~is the cavity roundtrip time, $\alpha$~is defined as half the total power loss per roundtrip, $\delta_0$ is the phase detuning between the driving field $E_\mathrm{in}$ and the closest cavity resonance, $\gamma$~is the nonlinear coefficient, and $\theta$~is the power coupling coefficient. Equation~\eqref{lle} includes dispersion to all orders, with $\beta_k$~the $k^\textrm{th}$~derivative of the propagation constant with respect to angular frequency.

To analyze the generation of sidebands via degenerate FWM, we consider the interaction between a strong CW field and two small-amplitude sidebands symmetrically detuned about the pump. The intracavity field can be approximated as $E(t,\tau)\approx A_0+a_{+1}\textrm{exp}(-\Omega \tau + \lambda t/t_\textrm{R})+a_{-1}\textrm{exp}(\Omega \tau + \lambda^* t/t_\textrm{R})$, where $A_0$ is a steady-state CW solution of Eq.~\eqref{lle}, $\Omega$ is the frequency detuning, and $a_{\pm1}$ are the sideband amplitudes at $t=0$ $(\lvert a_{\pm1}\rvert \ll \lvert A_0 \rvert)$. Injecting this ansatz into Eq.~\eqref{lle} yields, to first order in $a_\pm$, the following expression for the potentially unstable eigenvalue $\lambda$~\cite{haelterman92b}:
\begin{equation}
\begin{split}
\lambda(\Omega)=&-\alpha+iD_\textrm{o}(\Omega)L+\Bigl[\left(\gamma PL\right)^2\\ &-\left(2\gamma PL-\delta_0+D_\textrm{e}(\Omega)L\right)^2\Bigr]^{\frac{1}{2}}
\end{split}
\label{eigenvalue}
\end{equation}
where $D_\textrm{e}(\Omega)=\sum^\infty_{n=1}(\beta_{2n}\Omega^{2n})/(2n)!$ and $D_\textrm{o}(\Omega)=\sum^\infty_{n=1}(\beta_{2n+1}\Omega^{2n+1})/(2n+1)!$ represent even and odd orders of dispersion, respectively, and $P = |A_0|^2$.

Parametric oscillation can in principle occur at all frequency shifts $\Omega$ for which the real part of $\lambda$ is greater than zero. From Eq.~\eqref{eigenvalue}, it is straightforward to see that this condition is satisfied when $\gamma PL\leq\delta_0-D_\textrm{e}(\Omega)L\leq3\gamma PL$. When the sidebands grow from broadband noise, steady-state oscillation can be expected at the phase-matched frequency $\Omega_\mathrm{pm}$, which experiences the largest gain. This frequency satisfies the condition:
\begin{equation}
D_\textrm{e}(\Omega_\textrm{pm})L+2\gamma PL-\delta_0=0.
\label{phasematching}
\end{equation}
As in the case of singly-resonant fiber oscillators \cite{xu08,wong07}, we find that odd orders of dispersion play no role in setting the phase-matched oscillation frequency [see Eq.~\eqref{phasematching}]. For frequency shifts that are not too large, it is typically sufficient to truncate the Taylor expansion of the full dispersion curve at $\beta_4$, such that $D_\textrm{e}(\Omega)=\beta_2\Omega^2/2+\beta_4\Omega^4/24$ \cite{harvey03,pitois03,marhic04}. In this scenario, large frequency shift sidebands arise when operating with the pump in the normal dispersion regime ($\beta_2>0$), and in the presence of negative fourth-order dispersion ($\beta_4<0$). Furthermore, in this regime, small changes of the pump wavelength can translate into very large changes in phase-matched sideband frequencies, thus enabling wide tunability.

\begin{figure}[htb]
\centering
\includegraphics[width=\linewidth]{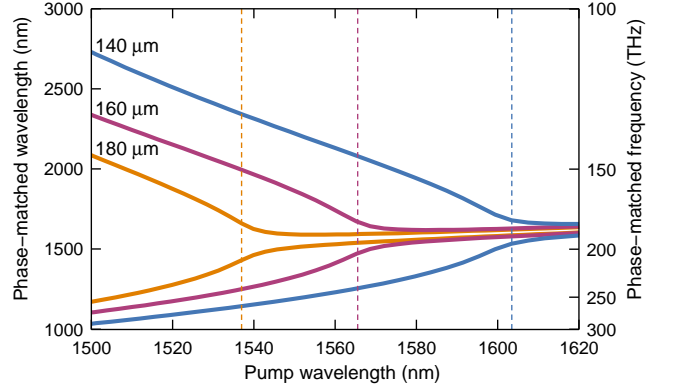}
\caption{Phase matching curves calculated from Eq.~\eqref{phasematching} for the fundamental TM mode of microspheres with diameters $140~\mu$m (orange), $160~\mu$m (magenta), and $180~\mu$m (blue). Dashed vertical lines indicate the corresponding ZDWs of the three spheres.}
\label{fig1}
\end{figure}

To illustrate the dependence of the phase-matched sidebands on the pump wavelength, we consider a silica microsphere similar to the experiments that will follow. It is well-known that the resonance wavelengths of a spherical dielectric cavity can be calculated directly from a characteristic equation~\cite{chiasera10,riesen15}. Accounting for the material dispersion of fused silica (and assuming the sphere to be surrounded by air), we can estimate the dispersion characteristics of the microspheres used in our experiments, and hence evaluate the phase-matched sideband positions via Eq.~\eqref{phasematching}. Figure~\ref{fig1} shows the phase-matched parametric sideband wavelengths as a function of pump wavelength for the fundamental TM mode family of three different sphere diameters (140, 160, and $180~\mu$m). For these calculations, we set the microsphere finesse ($\sim5\cdot10^4$), pump power (50~mW), nonlinear coefficient ($\gamma=2~$W$^{-1}$km$^{-1}$), and detuning ($\delta_0\sim0$) to values similar to the experiments that follow.

We first discuss the case of the $160~\mu$m-diameter sphere (solid magenta line in Fig.~\ref{fig1}). Also highlighted as the dashed magenta line is the position of the zero-dispersion wavelength (ZDW) of the fundamental TM mode at 1565.6~nm. As can be seen, for pump wavelengths above the ZDW, where the dispersion is anomalous, the negative value of $\beta_2$ yields small frequency shift sidebands with little tunability as the pump wavelength is varied. In stark contrast, for pump wavelengths below the ZDW, where the dispersion is normal, the phase-matched sidebands display much larger frequency shifts. This can be intuitively understood by noting that the positive contribution to $D_\textrm{e}(\Omega)$ from the $\beta_2\Omega^2$ term can only be compensated by the negative contribution of the $\beta_4\Omega^4$ term at large frequency shifts. Furthermore, it is evident from Fig.~\ref{fig1} that, when operating in the normal dispersion regime, small changes in the pump wavelength result in large variations of the phase-matched sideband wavelengths. As an example, a 60~nm change in pump wavelength in the normal dispersion regime, from 1500 nm to 1560 nm, is predicted to result in a combined tuning range in excess of an optical octave (from 1120 to 2270~nm) for the two parametric sidebands. Of course, dispersion engineering allows for the tuning range to be manipulated, and optimized for a given pump laser. For silica microspheres, this can be achieved simply by changing the sphere diameter, as shown in Fig.~\ref{fig1}.

\begin{figure}[t]
\centering
\includegraphics[width=\linewidth]{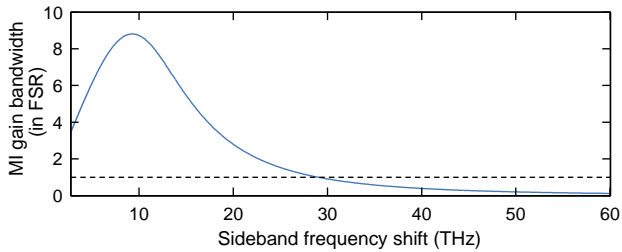}
\caption{Parametric gain bandwidth as a function of sideband frequency shift for a $160~\mu$m microsphere. This curve is calculated using Eq.~\eqref{eigenvalue} with the same parameters used in Fig.~\ref{fig1}. The gain bandwidth is larger (smaller) than one FSR above (below) the dashed line.}
\label{fig2}
\end{figure}

Equation~\eqref{eigenvalue} predicts the parametric gain envelope, but the actual frequency of oscillation must also be resonant with a mode of the cavity. As the parametric gain bandwidth (i.e., the range of frequencies for which $\textrm{Re}[\lambda(\Omega)]>0$) is non-zero, the conditions of optical gain and resonance can often be satisfied simultaneously. In Fig.~\ref{fig2}, we plot the parametric gain bandwidth as a function of the phase-matched frequency shift for a $160~\mu$m sphere using the same parameters as in Fig.~\ref{fig1}. For small sideband detunings, the gain bandwidth is large and spans over several free-spectral ranges (FSRs). In this situation, it is always possible to find a pair of cavity modes that will overlap with the gain spectrum. However, at larger detunings, the gain bandwidth reduces significantly, falling below one~FSR for shifts greater than around 30~THz. Accordingly, for such large frequency shifts, it may not always be possible to observe any parametric oscillation, as there may be no resonator modes that fall inside the parametric gain band. At even larger detunings the parametric gain bandwidth becomes smaller still, and precise dispersion engineering may be required to ensure that the phase-matched frequencies coincide with resonant modes of the cavity. Another important implication of the above discussion is that the tunability presented in this scheme is discrete (rather than continuous), as parametric oscillation is only possible when the sidebands are resonant with the cavity.

We now present our experimental demonstration of widely-tunable parametric oscillation in a silica microsphere. The sphere used in our experiments is formed by melting the end of an $80~\mu$m diameter silica optical fiber in a commercial fiber fusion splicer, and we measure its diameter and finesse to be approximately $163~\mu$m and $5\cdot10^4$, respectively. We drive the cavity with CW laser light derived from a C-band external cavity laser (New Focus Velocity TLB-6728, linewidth $\sim100$~kHz), whose output is amplified by an erbium-doped fiber amplifier and then spectrally filtered to remove unwanted amplified spontaneous emission noise. The resulting CW pump field has an average power of 50~mW, and it can be spectrally tuned from 1527~nm to 1565~nm. We couple the pump into the silica microsphere using an optical taper with a $1~\mu$m diameter waist. A polarization controller positioned before the taper allows the selection of the correct polarization for the fundamental TM mode family. At the output of the taper, the signal is split in two by a 50/50 optical coupler. Half of the signal is sent directly to an optical spectrum analyzer (OSA), while the other half passes through an offset bandpass filter (passband 1200 to 1400 nm) before detection by an amplified photodiode. 

To make conclusive observations of large frequency shift sidebands and to test their tunability, we employ the following experimental procedure. First, we coarsely tune the pump wavelength to a desired spectral region. We then look for resonator modes generating large frequency shift sidebands by scanning the pump wavelength over one free-spectral range while simultaneously monitoring the signal transmitted by the offset bandpass filter. Finally, we use fine piezo-tuning to tune the pump into resonance with a desired mode (from the short wavelength side), and thermally lock the pump-cavity detuning to a value where parametric oscillation is observed.

\begin{figure}[htb]
\centering
\includegraphics[width=\linewidth]{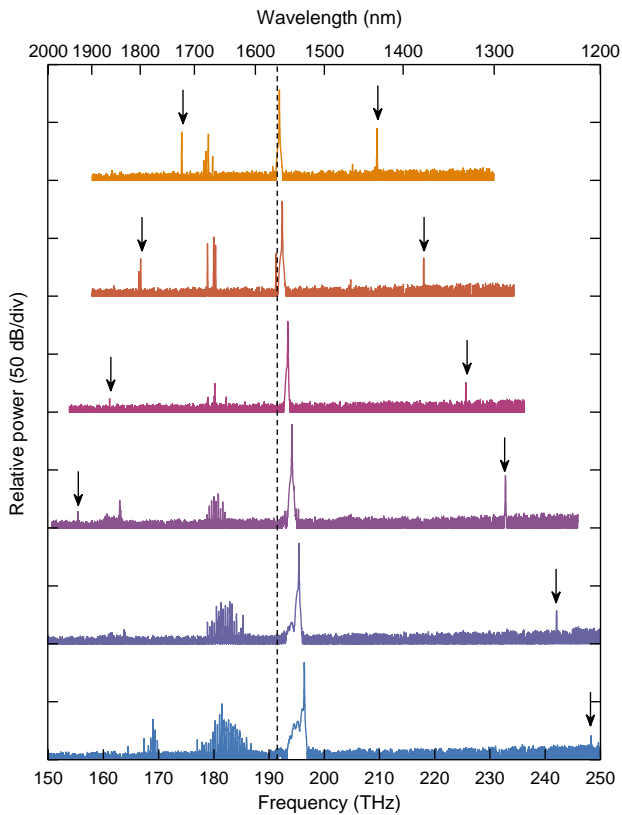}
\caption{Spectra of widely tunable sidebands observed in a $163~\mu$m diameter microsphere at pump wavelengths of (from top to bottom): 1563.7, 1559.5, 1551.1, 1545.3, 1535.3 and 1527.7~nm. The positions of the individual parametric sidebands are indicated by black arrows. The black dashed line indicates the position of the ZDW.}
\label{fig3}
\end{figure}

Figure~\ref{fig3} shows experimentally measured optical spectra at the microsphere output for six different pump wavelengths: 1563.7, 1559.5, 1551.1, 1545.3, 1535.3 and 1527.7~nm. Also shown is the phase-matched sideband wavelengths (black arrows) as well as the ZDW of the microsphere (dashed black line). From these measurements, we see clearly how the frequency detuning of the observed FWM sidebands increases as the pump wavelength is moved away from the ZDW  (deeper into the normal dispersion regime). Furthermore, the results demonstrate how small changes in the pump wavelength are magnified into large shifts of the FWM sidebands: even though the pump wavelength is only changed for about 35~nm, the combined tuning range of the sidebands exceeds 720~nm, stretching from 1207~nm to 1930~nm. We note that, for the two shortest pump wavelengths (1527.7 and 1535.3~nm), the long wavelength  FWM sideband falls below the noise floor of our OSA, and that more generally the sideband power levels display a clear asymmetry. We believe this asymmetry is simply due to the increased material absorption of fused silica at longer wavelengths, particularly above 1900~nm. Also visible in Fig.~\ref{fig3} is a broadband optical signal at a detuning of approximately $-13$~THz from the pump, which coincides with the peak gain of stimulated Raman scattering in fused silica \cite{stolen89,spillane02}. Provided this parasitic Raman frequency conversion process does not significantly deplete the intracavity pump field, it will have little effect on either the frequency shift or conversion efficiency of the FWM process.

\begin{figure}[htb]
\centering
\includegraphics[width=\linewidth]{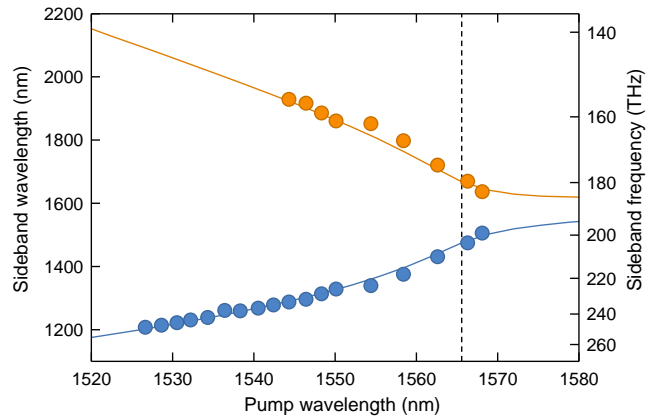}
\caption{Experimentally measured wavelengths of signal and idler waves as the pump wavelength is varied from 1568~nm to 1527~nm (solid circles). The solid line shows the theoretical phase matching curve [Eq.~\eqref{phasematching}] for a $160~\mu$m diameter silica microsphere. The black dashed line indicates the position of the ZDW.}
\label{fig4}
\end{figure}

A more comprehensive comparison between our experimental findings and theoretical phase matching predictions is shown in Fig.~\ref{fig4}. Here we plot the experimentally measured sideband wavelengths (solid circles) as a function of pump wavelength, as well as the phase matching curve predicted by Eq.~\eqref{phasematching} for the fundamental TM mode of a silica microsphere with a diameter of $160~\mu$m (solid curve). As can be seen, the experimentally measured sideband wavelengths agree very well with the theoretically predicted phase matching curve. This confirms that the widely-tunable parametric oscillation observed in our experiment indeed occurs when the pump is resonant with the fundamental TM mode family of the microsphere, and that the phase matching is due to higher-order dispersion.

In summary, we have presented the first experimental demonstration of widely tunable parametric oscillation in a Kerr nonlinear microresonator. Specifically, by pumping a silica microsphere in the normal dispersion regime with a CW laser tunable in the C-band, we have generated large frequency shift parametric sidebands whose combined tuning range extends from 1207 nm to 1930~nm. The (upper) wavelength limit in our current implementation is set by the strong material absorption of fused silica above $1.9~\mu$m; we expect that even larger tuning ranges can be reached by using crystalline resonators whose transparency window extends into the mid-IR. Such sources could then potentially allow access to the important 2--4~$\mu$m mid-IR region via direct conversion from a near-IR pump. Of course, as this work shows, for such extremely large frequency shifts, precise dispersion engineering may be necessary to ensure overlap between the parametric gain spectrum and resonant modes of the cavity. If successful, the resulting devices would offer the intriguing possibility of a compact and chip-scale parametric device with a widely tunable output, driven by a low-cost, low-power optical pump.

\textbf{Funding.} Marsden Fund and Rutherford Discovery Fellowships of the Royal Society of New Zealand.

%%%%%%%%%%%%%%%%%%%%%%%%%%%%%%%%%%%%%%%%%%%%%%%%%%%%%%%%%%%%%%%%%%%%%%%%%%%%%%%%%%%%%

\newcommand{\enquote}[1]{``#1''}

\end{document}